\documentclass[12pt]{iopart}

\expandafter\let\csname equation*\endcsname\relax
\expandafter\let\csname endequation*\endcsname\relax
\usepackage{amsmath}
\usepackage{amssymb}
\usepackage{graphicx}
\usepackage{cite}
\usepackage{bm}
\usepackage{array}
\usepackage{dcolumn}
\usepackage{hepparticles}
\usepackage{heppennames}
\usepackage{hepnicenames}
\usepackage{epstopdf}
\usepackage{xcolor}
\usepackage[page]{appendix}  

\newcommand{\ket}[1]{{| {#1} \rangle}}
\newcommand{\bra}[1]{{\langle {#1}|}}

\newcommand{\nbar}{\bar{n}}

\begin{document}

\title{Measurement-Induced Heating of a Trapped Ion}

\author{A.J. Rasmusson$^1$, Ilyoung Jung$^1$, Frank G. Schroer$^1$, Antonis Kyprianidis$^1$, and Philip Richerme$^{1,2,\dagger}$}
\address{$^1$Indiana University Department of Physics, Bloomington, Indiana 47405, USA}
\address{$^2$Indiana University Quantum Science and Engineering Center, Bloomington, Indiana 47405, USA}
\address{$^\dagger$richerme@indiana.edu}

\date{\today}

\begin{abstract}
We experimentally study the heating of trapped atomic ions during measurement of their internal qubit states. During measurement, ions are projected into one of two basis states and discriminated by their state-dependent fluorescence. We observe that ions in the fluorescing state rapidly scatter photons and heat at a rate of \mbox{$\dot{\bar{n}}\gtrsim 2\times 10^4$ quanta/s}, which is orders of magnitude faster than typical anomalous ion heating rates. We introduce a quantum trajectory-based framework that accurately reproduces the experimental results and provides a unified description of ion heating for both continuous and discrete sources.
\end{abstract}

\maketitle

\label{sec:intro}
\section{Introduction}
A trapped ion heats when it gains motional energy from its surrounding environment. Ion motion is the medium by which quantum information is transferred between qubits \cite{molmer1999multiparticle}, and motional heating is detrimental to high-fidelity quantum operations. For instance, heating is known to disrupt the ions' phase space trajectories during the application of entangling operations and necessarily leads to quantum gate errors \cite{haddadfarshi2016high, milne2020phase, valahu2022quantum}. Moreover, as ions heat, they become even more susceptible to errors arising from non-closure of phase space trajectories, noise in the driving fields, or motional frequency drifts \cite{bentley2020numeric,valahu2022quantum,cetina2022control}. Effects such as anomalous ion heating are so pernicious that they set constraints on ion trap designs \cite{deslauriers2006scaling,brownnutt2015ion,an2019distance} and entangling gate timings \cite{ballance2016high,gaebler2016high}, and they motivate specialized preparation of trap electrode surfaces \cite{allcock2011reduction,hite2012100} and operation of traps in cryogenic environments \cite{labaziewicz2008suppression,pagano2018cryogenic,weber2023cryogenic}.

Less explored are the heating effects from measurements in the middle of a quantum circuit, which are fundamental to multiple areas of quantum information processing. For instance, many quantum error-correcting protocols rely upon `mid-circuit' measurements and feedforward to correct errors and provide fault-tolerant operations \cite{chiaverini2004realization,terhal2015quantum, negnevitsky2018repeated,wan2019quantum,ryan2021realization, postler2024demonstration}. In addition, mid-circuit measurements are central to measurement-based quantum computing schemes \cite{raussendorf2003measurement, briegel2009measurement} and entanglement phase transitions in quantum many-body systems \cite{iqbal2024topological, iqbal2024non, skinner2019measurement,ippoliti2021entanglement,foss2023experimental}. They may also provide a more efficient way of encoding quantum algorithms on NISQ-era hardware \cite{chertkov2022holographic,decross2023qubit,moses2023race}.

Previous work in both trapped-ions and neutral atoms has noticed loss in gate fidelity, or atom loss, due to measurement-induced heating. Consequently, time-consuming recooling schemes were folded into experimental sequences to avoid such detrimental effects \cite{schindler2013undoing, pino2021demonstration, moses2023race, covey20192000,chow2023high}. However, the specific link between heating processes and mid-circuit measurements, which arise from fundamental photon-atom interactions, have yet to be fully characterized or quantified in trapped-ion systems. Consequently, how much recooling is required to guarantee a given level of error following a mid-circuit measurement was not predictable. Such an understanding would provide a verified model by which to evaluate recooling strategies and estimate potential errors after mid-circuit measurements due to heating.

Here, we experimentally study the measurement-induced heating of a trapped ion. We first establish the anomalous heating rate under ambient conditions, which is then compared to the observed heating rate during measurement of the qubit state. We find a measurement-induced heating rate of \mbox{$\gtrsim 2\times 10^4$ quanta/s}, which is $\sim 30$ times larger than the ambient heating in our trap. We develop a generalized theoretical framework to describe heating from continuous noise sources (leading to anomalous heating) and discrete noise sources (such as photon absorption and emission, responsible for heating during measurement). This framework supports our experimental observation that measurement-induced heating cannot be avoided by specific choices of detection laser parameters. We conclude that dedicated recooling strategies will be required for high-fidelity quantum operations following mid-circuit ion measurements.

\section{Ambient Heating}
Before characterizing the effects of measurement-induced heating, we first measure our baseline ambient heating conditions and introduce a predictive theoretical framework to describe heating in general. Motional state heating $\dot{\bar{n}}$ is defined as the rate at which the average phonon occupation $\bar{n} =  \langle \hat{a}^\dagger \hat{a} \rangle$ increases per unit time \cite{brownnutt2015ion}. Under ambient experimental conditions, $\nbar$ will increase due to the interaction of ions with various sources of external noise \cite{wineland1998experimental, brownnutt2015ion, hite2017measurements}, which generically may be modeled as a continuous, time-dependent fluctuating field \cite{lamoreaux1997thermalization}. 

Experiments are performed with a single $^{171}$Yb$^+$ ion confined in a four-rod linear Paul trap detailed in Ref. \cite{d2021radial} with radial secular frequency $\omega = 2 \pi \times 1.09$ MHz. Doppler cooling of the ions is accomplished by irradiating the 369.5 nm $^2$S$_{1/2}\ket{F=0}$ $\rightarrow$ $^2$P$_{1/2}\ket{F=1}$ and $^2$S$_{1/2}\ket{F=1}$ $\rightarrow$ $^2$P$_{1/2}\ket{F=0}$ transitions, with the dark qubit state $\ket{0} \equiv ^2$S$_{1/2}\ket{F=0}$ prepared via optical pumping. Far-detuned Raman beams at 355 nm drive carrier transitions between the hyperfine qubit states $\ket{0}$ and $\ket{1} \equiv {}^2$S$_{1/2}\ket{F=1 \; m_F=0}$ as well as red and blue sideband transitions \cite{wineland1998experimental}. The ion is cooled to near the motional ground state through pulsed resolved sideband cooling using second and first-order red sideband pulses \cite{rasmusson2021optimized}. The internal qubit states are detected by resonantly irradiating the $^2$S$_{1/2} \ket{F=1}$ $\rightarrow$ $^2$P$_{1/2}\ket{F=0}$ transition of the ion at 369.5 nm and collecting the state-dependent fluorescence on a photomultiplier tube (PMT) for 1 ms. The combined state preparation and measurement errors are estimated to be $<0.3\%$.

Our measurement of the ambient motional heating rate does not assume that the motional states follow a thermal distribution \cite{chen2017sympathetic, rasmusson2021optimized, reed2024comparison} and is executed as follows. The ion is first cooled to near the motional ground state. Then, the ion is left in the dark for a precise delay time gaining motional quanta due to ambient heating sources. A blue sideband is then driven from 0 - 300 $\mu$s (covering five periods) with 60 time points and $500$ repetitions per point. The median value for the low-energy motional state probabilities is then computed using the Singular Value Decomposition (SVD) method \cite{meekhof1996generation, turchette2000decoherence, rasmusson2021optimized} with Monte Carlo error propagation to estimate asymmetric $1\sigma$ confidence intervals. 

\begin{figure}[t]
    \centering
    \includegraphics[width=10cm]{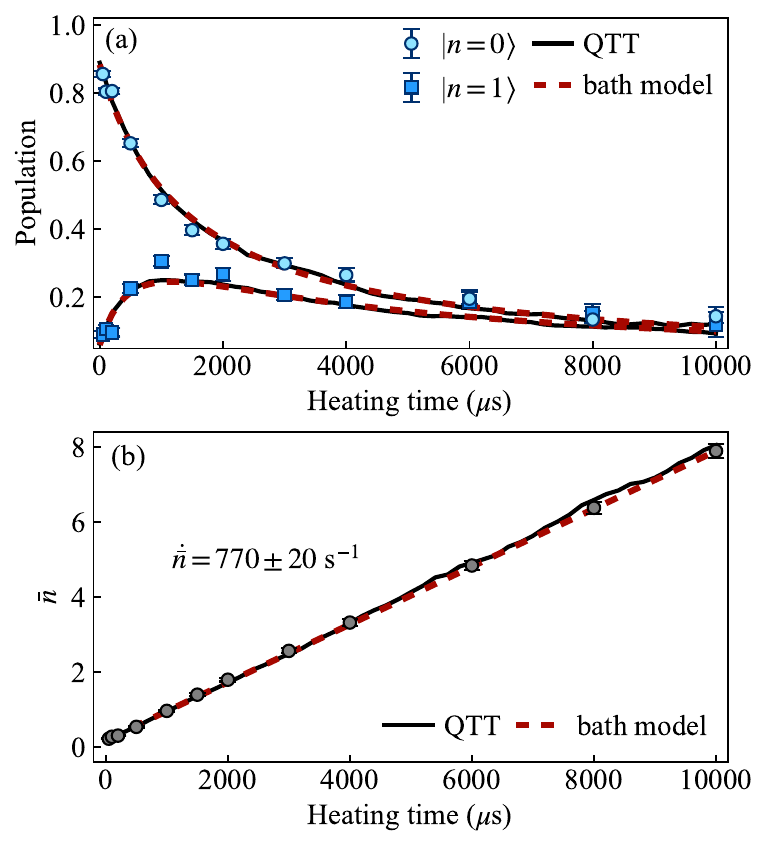}
    \caption{Motional state dynamics under ambient heating conditions. (a) Motional state probabilities are measured using the SVD method. Both a bath model \cite{turchette2000decoherence} (dashed red) and our Quantum Trajectory Theory (QTT) model (solid black) show good agreement with the experimental data. (b) The estimated $\nbar$ under ambient conditions shows linear heating and is well described by both models.}
    \label{fig:ambient-heating}
\end{figure}

As the delay time for ambient heating is varied, the motional state dynamics under ambient heating conditions are shown in Fig. \ref{fig:ambient-heating}(a). We plot the occupation probability for the lowest two motional energy levels $\ket{n=0}$ and $\ket{n=1}$ which have the largest dynamic range. Heating dynamics are observed as probability decays out of $\ket{n=0}$ and into $\ket{n=1}$, leading to a temporary increase of the $\ket{n=1}$ probability at early times. Afterwards, population in the states $\ket{n=0}$ and $\ket{n=1}$ monotonically decays towards thermal equilibrium with the environment.

The motional state dynamics may be modeled as an ion coupled to a reservoir, or `bath', outlined and experimentally observed in Ref. \cite{turchette2000decoherence}. The dynamics of the $n$th diagonal element of the motional state density matrix $\rho_{n,n}(t)$ follow the solution to the master equation for a harmonic oscillator weakly coupled to a high temperature reservoir---which is the case for ambient experimental conditions---and given by \cite{turchette2000decoherence}
\begin{align}
    \rho_{n,n}(t) &= \dfrac{1}{1 + \dot{\nbar} t} \sum_{j=0}^n \left(\frac{\dot{\nbar} t}{1 + \dot{\nbar}t}\right)^j \left(\frac{1}{1 + \dot{\nbar} t}\right)^{2n - 2j} \nonumber \\
    &\times \sum_{l=0}^\infty \left(\frac{\dot{\nbar}t}{1 + \dot{\nbar}t}\right)^l \begin{pmatrix} n + l -j \\ n - j \end{pmatrix} \nonumber \\
    &\times \begin{pmatrix} n \\ j \end{pmatrix} \rho_{n+l-j, n+l-j}(0)
\label{eq:rhonn-bath}
\end{align}
where $\dot{\nbar}$ is the linear heating rate and the only free parameter. Equation \eqref{eq:rhonn-bath} is fit to the measured motional state probabilities, yielding an ambient heating rate of $\dot{\nbar} = 770 \pm 20$ s$^{-1}$. In Fig. \ref{fig:ambient-heating}(b) we show the estimated $\nbar$ from a cumulative fit to the bath model (Appendix A). For this data, the initial state after sideband cooling is described by a double thermal distribution, estimated from the initial measured $\ket{n=0}$ and $\ket{n=1}$ values and extended to include the first 100 motional states \cite{chen2017sympathetic, rasmusson2021optimized}. Both the data and the bath model exhibit linear heating, as expected for trapped ion systems subject to continuous fluctuations of electric fields. 

\section{QTT Model}
We introduce a framework based on semi-classical quantum trajectory theory (QTT) \cite{carmichael2007statistical, horvath2007quantum} which can accurately predict motional state dynamics, $\nbar$, and $\dot{\nbar}$ for both continuous and discrete sources of ion heating. We first describe the approach, then establish its validity for continuous noise sources under ambient heating conditions by comparing it to our experimental data and to the bath model. 

Ambient heating sources are modeled as an effective fluctuating electric field $E(t)$ which captures a wide range of physical noise sources \cite{lamoreaux1997thermalization}.
To compute the quantum trajectory along the radial direction of interest, the classical center-of-mass phase space coordinate
\begin{equation}
    \alpha(t) = \sqrt{\frac{m\omega}{2 \hbar}}\hat{x}(t) + \frac{i}{\sqrt{2 m\omega\hbar}}\hat{p}(t)
\label{eq:alpha}
\end{equation}
with harmonic frequency $\omega$ and particle mass $m$ is recorded as $E(t)$ shifts the ion in phase space $\alpha \rightarrow \alpha + \alpha_k$. We consider a time interval $\Delta t = t_{k+1} - t_k$ that is long compared to the correlation time of the electric field fluctuations and satisfies $|\alpha_k| \ll 1$. For a fluctuating electric field, a particle of charge $e$ is shifted by \cite{james1998theory}
 \begin{equation}
     \alpha_k = \frac{i e}{\sqrt{2 m \omega \hbar}} \int_{t_k}^{t_{k+1}}E(t) e^{i \omega t} dt.
 \label{eq:alpha_k}
 \end{equation}
We then compute the motional state dynamics from the displacement $\hat{D}(\alpha)$ of the initial motional state $\ket{n}$ to Fock state $\ket{m}$. For $m \geq n$,
\begin{align}
    p_{m}(n) &= |\bra{m}\hat{D}(\alpha) \ket{n}|^2 \nonumber \\
    & = \frac{n!}{m!} |\alpha|^{2(m-n)} e^{-|\alpha|^2} \bigg[ \mathcal{L}_n^{(m-n)}(|\alpha|^2) \bigg]^2
\label{eq:displaced-fock-state}
\end{align}
where $\mathcal{L}_n^{(k)}(x)$ is the generalized Laguerre polynomial \cite{de1990properties}. Combining this with the definition of $\alpha_k$ in Eq. \eqref{eq:alpha_k} (see Appendix B), we recover the commonly defined heating rate in the literature \cite{wineland1998experimental,brownnutt2015ion}, $\dot{\nbar}=e^2 S_E(\omega)/(4 m \omega \hbar)$, where $S_E(\omega)$ is the spectral density of electric field fluctuations at the trap frequency. Quantum trajectories randomly sample electric fields, using a $S_E(\omega)$ independently determined by the bath model, and are averaged together to provide the expected motional heating. On average, the microscopic phase space kicks $\alpha_k$ are connected to the macroscopic observable $\nbar$ since $\langle|\alpha_k|^2\rangle = \dot{\nbar} \Delta t$.

Figure \ref{fig:ambient-heating} shows that the QTT model, averaged over 1000 trajectories, closely agrees with the experimental data and the bath model in predicting motional state dynamics and $\nbar$. We emphasize that our QTT model contains no adjustable parameters and depends only on the physical quantities in Eq.~\eqref{eq:alpha_k}. In the next section, we will show how QTT may be readily adapted to discrete heating sources, such as photon kicks, by adjusting the definition of $\alpha_k$ accordingly.


\section{Measurement-Induced Heating}
We now consider the heating of a trapped ion irradiated with a resonant detection beam, which is the standard configuration for qubit state readout. Two experiments are performed: one where an ion is prepared in the dark qubit state ($\ket{0}$), and a second where an ion is prepared in the bright qubit state ($\ket{1}$). In each case, the ion is first cooled to near its motional ground state by pulsed sideband cooling \cite{rasmusson2021optimized}. If a bright qubit state is desired, we drive a carrier $\pi$-pulse from $\ket{0}$ to $\ket{1}$ with measured fidelity 99.3 $\pm$ 0.1 \%. The detection beam is then turned on for a variable heating time.
An optical pumping pulse then resets the qubit state to $\ket{0}$ while preserving the newly excited phonon state $\ket{n}$. Finally, the phonon state probabilities are measured using the same SVD method outlined in the previous section.

\begin{figure}[t]
    \centering
    \includegraphics[width=10cm]{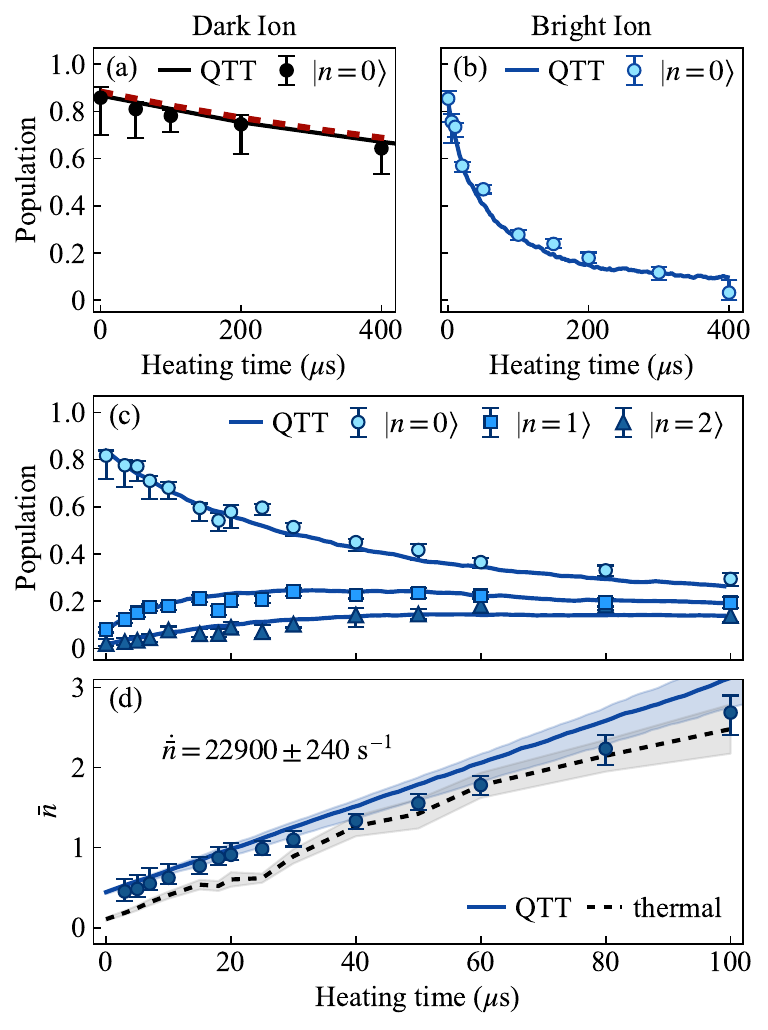}
    \caption{Motional state dynamics during measurement-induced heating. (a) Motional ground state $\ket{n=0}$ probability for a dark ion. The bath model (dashed red) under the ambient conditions of Fig.~\ref{fig:ambient-heating}(a) is included for comparison. (b) The motional ground state probability for bright-state ions decays much more rapidly than for dark-state ions. (c) Early-time population dynamics of a bright ion undergoing measurement-induced heating. The QTT model accurately describes the data with no adjustable parameters. (d) Estimates of $\nbar$ based on the panel (c) data, comparing the QTT model to a thermal model. Shaded bands indicate $1\sigma$ model uncertainties. Errors in the QTT model arise from experimental measurement errors in the scattering rate $\Gamma$.}
    \label{fig:fig2}
\vspace{-6px}
\end{figure}

We expect that ions in the dark state will scatter no photons during measurement and exhibit the same heating rate as the ambient case. This expectation is confirmed in Fig. \ref{fig:fig2}(a), which shows the probability of $\ket{n=0}$ for dark ions during measurement. Data were taken out to 8 ms and fit to a global bath model, yielding an estimated motional heating rate of $780 \pm 40$ s$^{-1}$. This rate is indistinguishable from the ambient case presented in Fig. \ref{fig:ambient-heating}(b) ($770 \pm 20$ s$^{-1}$).

In stark contrast, ions in the bright qubit state will undergo stochastic momentum kicks of order $\sim \hbar k$ due to photon absorption and emission. This process cannot be generically described by the bath model of Eq.~\eqref{eq:rhonn-bath}. Although photon scattering does not increase the average velocity of the ion ($\langle v \rangle = 0$ in a trap), the stochastic timing of photon recoil events relative to the harmonic motion of the ion is expected to increase velocity fluctuations $\langle v^2 \rangle$ and therefore increase motional energy \cite{itano1982laser}. This mechanism is observed in Fig.~\ref{fig:fig2}(b) where the $\ket{n=0}$ probability of a prepared bright ion decays rapidly, indicating fast heating out of the motional ground state.

We model this process using our QTT framework, where the phase-space kick $\alpha_k$ now encodes the effects of atomic scattering processes. We write (Appendix C):
\begin{align}
\label{eq:alpha-k-detection}
    \alpha_k = \frac{i e^{i \omega t} \hbar k }{\sqrt{2m\omega\hbar}}\bigg[\sqrt{f_x} + \sqrt{f^{(k)}_{sx}}
 + \frac{8 \sqrt{f_x} \Delta \omega n}{\gamma^2(1+s')+4\Delta^2} \bigg] \;
\end{align}
where $\omega = 2\pi \times 1.09$ MHz is the trap frequency along the radial $x-$axis, $k = 2\pi / (369.5 \; \text{nm})$ is the wave vector, and $n$ is the ion motional state. In Eq. \eqref{eq:alpha-k-detection}, absorption is described by the geometric factor $f_x = 1/4$ which accounts for the projection of the incident laser beam along the $x-$axis. Emission is treated with geometric factor $f^{(k)}_{sx}$, randomly chosen to recreate an isotropic emission pattern \cite{itano1982laser,horvath2007quantum}. Finally, the Doppler effect is included as the final term in Eq.~\eqref{eq:alpha-k-detection} and depends on parameters such as the natural linewidth $\gamma$, modified saturation parameter $s'$, and laser detuning $\Delta$ \cite{ejtemaee2010optimization}.

The overall measurement-induced heating rate is given by the average energy gained per scattering event, $\hbar\omega |\alpha_k|^2$, times the scattering rate $\Gamma$. We determine $\Gamma$ by fitting a measured fluorescence versus intensity curve to the scattering rate equation appropriate for $^{171}$Yb$^+$ \cite{ejtemaee2010optimization}
\begin{equation}
    \Gamma = \frac{\gamma (s / 18)}{1 + \frac{1}{216}(\frac{s \gamma}{\delta_B})^2 + \frac{8}{3}(\frac{\delta_B}{\gamma})^2 + (\frac{2 \Delta}{\gamma})^2}
\label{eq:scat-rate}
\end{equation}
where $s \equiv I/I_\text{sat}$ is the saturation parameter for the detection transition at 369 nm, $\Delta$ is the laser detuning, $\gamma = 2 \pi \times 19.6$ MHz is the natural linewidth, and $\delta_B = 2\pi \times 5.288$ MHz is the Zeeman splitting. For the saturation parameter $s = 1.27 \pm 0.02$ used in these experiments, the corresponding scattering rate is $\Gamma = 2\pi\times (1.07 \pm 0.01) $ MHz, with the error arising predominantly from uncertainty in $s$. Consequently, the QTT model inherits this uncertainty in $s$ and $\Gamma$ when predicting the motional heating rate $\dot{\bar{n}}$.


We report agreement between the QTT model and the experimental data to within error bars in Fig.~\ref{fig:fig2}.
The model accurately predicts the dynamics of the motional ground state (Fig.~\ref{fig:fig2}(a)-(b)) and all measured low-lying motional states (Fig.~\ref{fig:fig2}(c)). This agreement at short times is particularly noteworthy, since most heating models assume thermal state distributions and would fail to capture the initial non-thermal behavior following sideband cooling \cite{chen2017sympathetic, rasmusson2021optimized, reed2024comparison}. As before, the QTT model in Fig. \ref{fig:fig2} contains no adjustable parameters and depends only on the independently-measured scattering rate, atomic physics, and the laser properties described in Eq. \eqref{eq:alpha-k-detection}.

Figure \ref{fig:fig2}(d) shows the rapid increase in $\nbar$ during measurement-induced heating. The QTT prediction (dark blue line) shows a linear heating rate of $\dot{\nbar} = 22900 \pm 240$ s$^{-1}$, which is nearly $30$ times faster than the heating rate for ions in the dark state or under ambient heating conditions. For comparison, we also fit the measured low-energy states $\ket{n=0}$, $\ket{n=1}$, and $\ket{n=2}$ to a presumed thermal distribution at each time point (dashed black line). The thermal fit significantly underestimates $\nbar$ at early times, consistent with previous numeric simulations and experimental observations \cite{chen2017sympathetic, rasmusson2021optimized,reed2024comparison} and only matches the experimental data after several hundred scattering events have taken place.

Since the typical qubit measurement time for trapped ions is significantly longer than the $100~\mu$s timescale shown in Fig. \ref{fig:fig2}(c)-(d), we investigate whether this rapid linear heating persists at longer times. As seen in Eq. \eqref{eq:alpha-k-detection}, the choice of laser detuning $\Delta$ plays a key role in determining the typical magnitude of phase-space kicks during photon scattering. Therefore, we study measurement-induced heating in the long-time limit for three different choices of $\Delta$, spanning the red-detuned, near-resonant, and blue-detuned regimes.

When the average ion energy exceeds $\nbar\approx 10$, it becomes difficult for traditional sideband thermometry techniques to accurately measure the motional state \cite{diedrich1989laser,wineland1998experimental,rasmusson2021optimized}. For this reason, we use a carrier Rabi oscillation to estimate $\nbar$ after extended heating times of up to 2 ms. Following measurement-induced heating of the trapped ion, a carrier oscillation is driven from 0 - 40 $\mu$s (covering 6 periods) with 60 time points and 300 repetitions each. The carrier oscillation is fit with $\bar{n}$ as a free parameter, taking into account all appropriate Debye-Waller factors (see Appendix E) \cite{wineland1998experimental}. This fitting procedure assumes a thermal motional state distribution, which is in agreement with our data after $\sim50~\mu$s (Fig. \ref{fig:fig2}(d)). 

\begin{figure}
    \centering
    \includegraphics[width=8.4cm]{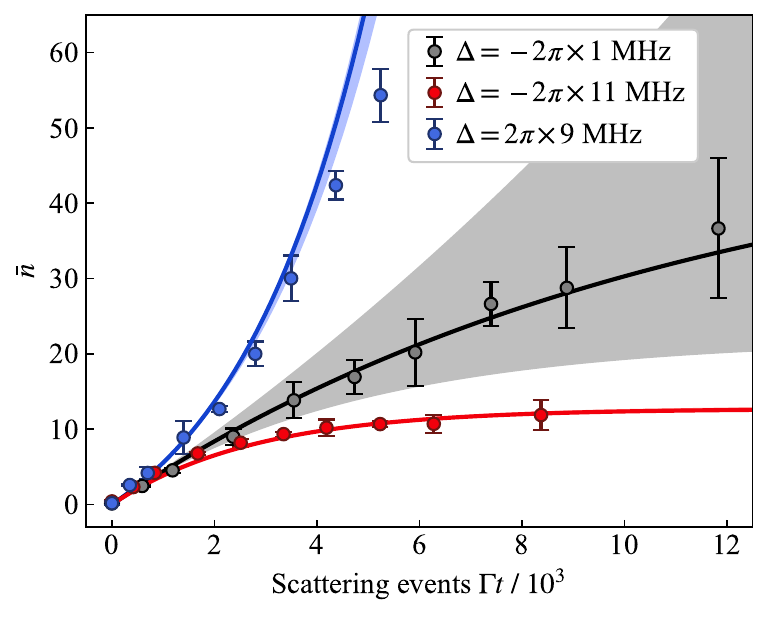}
    \caption{Measurement-induced heating depends sensitively on the detuning $\Delta$ of the detection beam. Darker solid lines are predicted values according to Eq. \eqref{eq:atomic-energy} with shaded areas showing $\pm$ 2 MHz uncertainty bands in detuning. Points and lines of the same color have the same detuning.}
    \label{fig:long-times}
\end{figure}

Figure \ref{fig:long-times} shows the sensitivity of ion motional heating to laser detuning $\Delta$ in the long-time limit. We perform the experiment with three different detection beam detunings: $\Delta = 2\pi \times \{-11, -1, 9\}$ MHz, corresponding to red-detuned, near-resonant, and blue-detuned, respectively. In Fig. \ref{fig:long-times}, the $x$-axis is scaled by the photon scattering rate of each detuning, $\Gamma = 2\pi \times \{0.56, 0.94, 0.67\}$ MHz as calculated from the scattering rate for $^{171}$Yb$^+$ provided in Eq. \ref{eq:scat-rate} \cite{ejtemaee2010optimization}. Our typical state detection requires \mbox{$\Gamma t \approx 5\times 10^3$} scattering events. For the red-detuned case, where $\Delta$ is chosen near the optimal Doppler detuning, the data equilibrates to the Doppler cooling limit $\nbar \approx 12.7$. All other choices for $\Delta$ leave the ion with higher motional energy in the long-time limit.

The expected $\nbar(t)$ under these conditions may be predicted by extending the QTT model with Eq. \eqref{eq:alpha-k-detection} to long times, or by adapting the equations governing laser cooling \cite{itano1982laser} to describe measurement-induced heating. We find (Appendix D):
\begin{equation}
    \nbar(t) = \left[\left(\nbar(0) - \frac{R}{D\hbar\omega}\right)e^{-\Gamma_0 D t} + \frac{R}{D\hbar\omega}\right]
\label{eq:atomic-energy}
\end{equation}
where we define $\Gamma_0 \equiv \gamma (s/18)/(1 + s^{\prime} + 4\Delta^2/\gamma^2)$ as the effective scattering rate, $R \equiv (f_x + f_{sx}) \hbar^2 k^2/(2m)$ as the effective recoil energy, and $D \equiv 8\Delta\hbar f_x k^2/(m\gamma^2(1+s^{\prime})+4m\Delta^2)$ as a Doppler-effect term that takes into account the laser detuning $\Delta$. In the expressions above, we assume that spontaneous emission is isotropic ($f_{sx} = 1/3$) and that the initial Doppler shift $\vec{k}\cdot\vec{v}$ is small compared to the natural linewidth $\gamma$. The parameters in Eq. \eqref{eq:atomic-energy} are independently measured, and the predicted curves for each detuning are plotted as solid lines in Fig. \ref{fig:long-times}. Shaded bands illustrate our $\pm 2$ MHz experimental uncertainty in measuring $\Delta$.

From this model and the data in Fig. \ref{fig:long-times}, we find that rapid measurement-induced heating cannot be avoided by judicious choice of detection beam parameters. At the start of a measurement, the short-time behavior of Eq.~\eqref{eq:atomic-energy} may be approximated as $\nbar(t) \approx \nbar(0)+\Gamma_0 R t/(\hbar\omega)$. This corresponds to linear initial heating at rate $\dot{\bar{n}}=\Gamma_0 R/(\hbar\omega) \sim2\times10^4$ s$^{-1}$, in agreement with measurements in Fig. \ref{fig:fig2}(d). The only tunable beam parameter is the photon scattering rate $\Gamma_0$, which cannot be reduced without sacrificing the qubit state detection fidelity or increasing the detection time (leading to the same final $\nbar$). We also note that this measurement-induced heating rate in $^{171}$Yb$^+$ is among the \emph{smallest} for common trapped-ion species, since the effective recoil energy $R$, and hence $\dot{\bar{n}}$, are suppressed by the large ion mass. 

\emph{Discussion and Outlook}---Mid-circuit measurements will be a critical element to future quantum technologies. This work has quantified the rapid motional heating which takes place during the measurement of a trapped ion, which dominates over anomalous heating and is an unavoidable consequence of photon scattering. We have also provided a unified framework (QTT) to accurately model motional heating effects for both continuous and discrete noise sources. Our experimental data and theoretical modeling indicate that, if left unmitigated, measurement-induced heating will present a substantial roadblock for performing high-fidelity operations following a mid-circuit measurement.

The effects of measurement-induced heating become more complex when detecting the state of specific ions in a larger array. For a collection of co-trapped ions, measurement-induced heating will add motional energy to all vibrational modes in which a bright ion participates, weighted by their mode participation amplitudes. Thus, measurement of a single ion within an array may demand recooling of multiple vibrational modes before high-fidelity gate operations may proceed.

We conclude that for mid-circuit measurements to be viable, dedicated mid-circuit recooling strategies must be implemented to combat measurement-induced heating. Brute-force sympathetic cooling has been successful at reducing multiple sources of ion heating, at the cost of co-trapping multiple ion species and spending more time recooling than executing quantum gate operations \cite{pino2021demonstration, moses2023race}. More sophisticated recooling schemes may be needed, such as rapid exchange cooling \cite{fallek2024rapid}, phonon rapid adiabatic passage cooling \cite{fabrikant2024cooling}, or perhaps dedicated mid-circuit measurement zones within a QCCD trap \cite{kielpinski2002architecture,pino2021demonstration}. Future work may also consider simultaneous ion measurement and recooling by utilizing shelving, state teleportation \cite{schmidt2005spectroscopy}, or the \textit{omg} qubit architecture \cite{allcock2021omg}.

\section*{Acknowledgements}
This material was based on work supported by the U.S. National Science Foundation, under Grant CHE-2311165. The IUB Quantum Science and Engineering Center is supported by the Office of the Vice Provost for Research through its EAR program.

\section*{References}
\bibliographystyle{unsrt}
\bibliography{main}{}

\begin{appendices}

\section{Determination of $\bar{n}$ from the Bath Model}
To estimate $\bar{n}$, we describe the initial motional state by a double thermal distribution, estimated from the initial measured $\ket{n=0}$ and $\ket{n=1}$ values and extended to include the first 100 motional states \cite{chen2017sympathetic, rasmusson2021optimized}. We then compute multiple fits of the motional heating rate $\dot{\bar{n}}$ to the data using the bath model. A cumulative fit is used for each point, such that $\dot{\bar{n}}_k$ determined at the $k$th time point only includes data up to time $t_k$. Each heating rate $\dot{\bar{n}}_k$ is then used to propagate the initial motional state following the bath model dynamics, resulting in an estimate of $\bar{n}$ at time $t_k$.

\section{Derivation of Continuous Heating Rate from QTT Model}

To compute the semi-classical quantum trajectories, the classical center-of-mass phase space coordinate
\begin{equation}
    \alpha(t) = \sqrt{\frac{m\omega}{2 \hbar}}\hat{x}(t) + \frac{i}{\sqrt{2 m\omega\hbar}}\hat{p}(t)
\end{equation}
is recorded as external sources stochastically shift $\alpha \rightarrow \alpha + \alpha_k$ over a small time interval $\Delta t = t_{k+1} - t_k$, such that $\alpha_k \ll 1$ \cite{carmichael2007statistical, horvath2007quantum}. Ambient heating sources are modeled as an effective fluctuating electric field $E(t)$, which captures a wide range of physical noise sources \cite{lamoreaux1997thermalization}. For a fluctuating electric field, a particle of charge $e$ is shifted by \cite{james1998theory}
 \begin{equation}
     \alpha_k = \frac{i e}{\sqrt{2 m \omega \hbar}} \int_{t_k}^{t_{k+1}}E(t^\prime) e^{i \omega t^\prime} dt^{\prime}.
 \label{eq:alpha_k2}
 \end{equation}
As stated in the main text, we then compute the motional state dynamics from the displacement $\hat{D}(\alpha)$ of the initial motional state $\ket{n}$ to Fock state $\ket{m}$ with $m \geq n$:
\begin{align}
    p_{m}(n) &= |\bra{m}\hat{D}(\alpha) \ket{n}|^2 \nonumber \\
    & = \frac{n!}{m!} |\alpha|^{2(m-n)} e^{-|\alpha|^2} \bigg[ \mathcal{L}_n^{(m-n)}(|\alpha|^2) \bigg]^2 \;.
\label{eq:displaced-fock-stateAP}
\end{align}

We observe that this probability depends on the square of the phase space displacement, $|\alpha|^2$. From our definition of $\alpha_k$, the square of the phase space kick is:
    \begin{equation}
        |\alpha_k|^2 = \frac{e^2}{2m\hbar\omega}\left|\int_{t_k}^{t_{k+1}} E(t') e^{i\omega t'} dt'\right|^2 \;.
    \end{equation}

Following Savard \cite{savard1997laser}, and using the Wiener-Khinchin theorem for time averages where the autocorrelation function decays to zero for long times, the squared integral on the right-hand side may be replaced by an integral over the autocorrelation of the electric field fluctuations:
\begin{equation}
     |\alpha_k|^2=\frac{e^2}{2m\hbar\omega} \Delta t \int_{-\infty}^\infty d\tau e^{i \omega \tau} \langle E(t) E(t+\tau) \rangle \;.
     \label{eq5}
\end{equation}
Finally, under the definition of the single-sided spectral density of electric field fluctuations \cite{turchette2000heating},
\begin{equation}
    S_E(\omega) = 2 \int_{-\infty}^\infty d\tau e^{i \omega \tau} \langle E(t) E(t+\tau) \rangle
\end{equation}
we can rewrite Eq. \eqref{eq5} in terms of the spectral density of electric field noise:
    \begin{equation}
        |\alpha_k|^2 = \frac{e^2}{4m\hbar\omega}S_E(\omega)\Delta t = \dot{\bar{n}} \Delta t \;.
    \end{equation}
    
\section{Derivation of Discrete Heating Rate from QTT Model}

Here, we calculate the change in phase space coordinate due to discrete photon scattering events. Near-resonance incident photons apply an average force on $^{171}$Yb$^+$ in the $x-$direction, quoted from Ref. \cite{itano1982laser} as
\begin{equation}
    \langle F_x \rangle = \Gamma_0 \hbar k \sqrt{f_x}\left(1 + \frac{8 \Delta n \omega }{\gamma^2(1+s')+4\Delta^2}\right)
\end{equation}
with variable definitions matching those given the text. The scattering rate appropriate for $^{171}$Yb$^+$ is \cite{ejtemaee2010optimization}
\begin{equation}
    \Gamma_0 = \frac{\gamma (s/18)}{1+s'+\frac{4\Delta^2}{\gamma^2}} \;.
\end{equation}
where the natural line width $\gamma = 2\pi \times 19.6$ MHz, and
\begin{equation}
    s^{\prime} = \frac{1}{216}\left(\frac{s \gamma}{\delta_B}\right)^2 + \frac{8}{3}\left(\frac{\delta_B}{\gamma}\right)^2
\end{equation}
where in our experiments the Zeeman splitting is $\delta_B = 2\pi \times 5.288$ MHz and the saturation parameter $s = 1.27$.

The momentum kick per absorption event $\Delta p_{\text{abs}}$ includes the Doppler related frequency shifts
\begin{equation}
    \Delta p_{\text{abs}} = \hbar k \sqrt{f_x}\left(1 + \frac{8 \Delta n \omega}{\gamma^2(1+s')+4\Delta^2}\right) \;.
\end{equation}

Momentum kicks due to emission are independent of velocity, delivering a momentum change
\begin{equation}
    \Delta p_{\text{em}} = \hbar k \sin(\theta) \cos(\phi)
\end{equation}
with angles $\theta$ and $\phi$ randomly chosen to generate an isotropic emission pattern \cite{itano1982laser}.


The change in phase space $\alpha_k$ due to a photon scattering event (absorption and emission) then is calculated as
\begin{align}
    \alpha_k =& \frac{i e^{i \omega t}}{\sqrt{2m\omega\hbar}}\Delta p \nonumber \\
    \alpha_k =& \frac{i e^{i \omega t} \hbar k}{\sqrt{2m\omega\hbar}}\bigg[\sqrt{f_x}\left(1 + \frac{8 \Delta n \omega}{\gamma^2(1+s')+4\Delta^2}\right) \nonumber \\
    &+\sin(\theta)\cos(\phi)  \bigg]
\end{align}
where the change in $\alpha$ due to absorption events has already been averaged over while the change in $\alpha$ due to emission will be averaged over in the quantum trajectory Monte Carlo. A single scattering event is on the order of 10 ns while the trap period is 1 $\mu$s. Therefore, we have assumed the same phase for an absorption-emission event pair.

This microscopic calculation of $\alpha_k$ generates the exact same heating dynamics $\dot{\bar{n}}$ as the more common semi-classical Doppler shifted energy equations \cite{itano1982laser} as will be shown below. The average motional energy $E = \hbar \omega |\alpha_k|^2$ is given by
\begin{equation}
    E = \frac{\hbar^2 k^2}{2m}\left[f_x \left(1 + \frac{8 \Delta n \omega}{\gamma^2(1+s')+4\Delta^2}\right)^2 + f_{sx}\right]
\label{eq:average-motional-energy}
\end{equation}
where the isotropic emission geometry factor $f_{sx} = \langle \sin(\theta)^2\cos(\phi^2)\rangle = 1/3$. Applying the binomial approximation, the average motional energy $E$ of Eq. \eqref{eq:average-motional-energy} recovers the result from the semi-classical Doppler derivation.

\begin{equation}
    E = \left[\frac{\hbar^2 k^2}{2m} (f_x + f_{sx}) + \frac{8\Delta \hbar f_x k^2/m}{\gamma^2(1+s')+4\Delta^2}n\hbar\omega \right] \;.
\end{equation}
This approach is in complete agreement with the semi-classical analysis. The energy change per scattering event is given by $\hbar \omega |\alpha_k|^2$, and the total rate of energy change $dE/dt = \Gamma_0 \hbar \omega |\alpha_k|^2$. Comparing this expression with the semi-classical calculation for $dE/dt$ (Eq. \eqref{eq:dexdt4} below), we arrive at the same result having started from a microscopic phase space displacement picture.

\section{Semi-classical Laser Heating}

We consider the regime where a near-resonant laser irradiating a harmonically trapped atom can cause heating of the ion's motion. Our treatment initially follows Ref. \cite{itano1982laser} but then deviates to consider the case of heating due to the laser frequency being near-resonance or blue-detuned ($\Delta > 0$).

Consider a resonant beam of intensity $I$ and wavevector $\vec{k}$. When this light hits a trapped ion moving with velocity $\vec{v}$ and a photon is absorbed, a photon is later emitted with wavevector $\vec{k}_s$.


To find the energy rate of change along a direction of interest (i.e. $x$-axis), we multiply the change in energy by the scattering rate $\Gamma(\omega, \vec{v})$ and average over the absorption and emission scattering directions:
\begin{equation}
\label{eq:dexdt}
    \frac{dE_x}{dt}=\langle \Gamma(\omega,\vec{v})\left[\frac{\hbar^2 k^2}{2m}(f_x+f_{sx})+\hbar k_x v_x\right] \rangle
\end{equation}
where $m$ is the mass of the trapped ion, $v_x$ is the velocity in along the $x$-axis, $k$ is the magnitude of $\vec{k}$, $f_x = \hat{k}_x^2$ is the incident photon geometric factor along the $x$-direction, and $f_{sx} = \hat{k}_{sx}^2$ is the emitted photon geometric factor along the $x$-direction.

The scattering rate $\Gamma(\omega,\vec{v})$ from Refs. \cite{ejtemaee2010optimization, berkeland2002destabilization} is given by:
\begin{equation}
\label{eq:gamma1}
    \Gamma(\omega,\vec{v})=\frac{\gamma (s/18)}{1+s^{\prime}+\frac{4(\Delta+\vec{k}\cdot\vec{v})^2}{\gamma^2}}
\end{equation}
where $\gamma=2\pi \times 19.6$~MHz is the natural linewidth, $s=I/I_{sat}$ is the saturation parameter, $\Delta$ is the detuning from resonance, and 
\begin{equation}
    s^{\prime} = \frac{1}{216}\left(\frac{s \gamma}{\delta_B}\right)^2 + \frac{8}{3}\left(\frac{\delta_B}{\gamma}\right)^2
\end{equation}
with Zeeman splitting $\delta_B$.

The Doppler shift $\vec{k}\cdot\vec{v}$ is much smaller than the natural linewidth such that Eq. \eqref{eq:gamma1} may be approximated by
\begin{equation}
\label{gamma2}
    \Gamma(\omega,\vec{v}) \approx \Gamma_0\left[1 + \frac{8\Delta (\vec{k}\cdot\vec{v})}{\gamma^2(1+s^{\prime})+4\Delta^2}\right]
\end{equation}
where
\begin{equation}
    \Gamma_0 = \frac{\gamma (s/18)}{1+s^{\prime}+\frac{4\Delta^2}{\gamma^2}} \;.
\end{equation}

The change in energy (Eq. \eqref{eq:dexdt}), assuming $\langle v_i \rangle =0$ and $\langle v_i v_j \rangle =0$ for any two directions $i\neq j$, becomes
\begin{equation}
\label{eq:dexdt4}
    \frac{dE_x}{dt}= \Gamma_0\left[ \frac{\hbar^2 k^2}{2m}(f_x + f_{sx}) + \frac{8\Delta\hbar f_x k^2 E_x /m}{\gamma^2(1+s^{\prime})+4\Delta^2} \right]
\end{equation}
where the average classical harmonic oscillator energy $E_x = m\langle v_x^2 \rangle$ has been swapped in (with no factor of 1/2).

Equation \eqref{eq:dexdt4} is a first order differential equation which can be solved analytically. For convenience, we define:
\begin{align}
    R & \equiv \frac{\hbar^2 k^2}{2m}(f_x + f_{sx}) \\
    D & \equiv \frac{8\Delta\hbar f_x k^2/m}{\gamma^2(1+s^{\prime})+4\Delta^2}
\end{align}
then our equation may be written
\begin{equation}
\label{eq:ABCeqn}
    \frac{dE_x}{dt}=\Gamma_0(R - D E_x)
\end{equation}
which has the solution
\begin{equation}
\label{eq:exsoln}
    E_x(t) = \left(E_x(0)-\frac{R}{D}\right)e^{-\Gamma_0 D t} + \frac{R}{D} \;.
\end{equation}
Replacing the energy with the average motional state $\bar{n} = E_x(t) / (\hbar \omega)$, we arrive at equation in the main text.

We examine Eq. \eqref{eq:exsoln} in a few limits. First, in the long-time limit, the final value of the energy converges to:
\begin{equation}
    E_x(\infty) = \frac{R}{D} = \frac{\hbar \gamma}{8}\left(1+\frac{f_{sx}}{f_x}\right)\left[\frac{\gamma(1+s^{\prime})}{2\Delta}+\frac{2\Delta}{\gamma}\right] \;.
\end{equation}
This agrees with results in the literature \cite{leibfried2003quantum}, which are typically derived by setting $dE_x/dt = 0$ in Eq. \eqref{eq:dexdt4}.

Another limit is the special case of zero detuning ($D=0$). In the limit of $D\rightarrow 0$, the motional energy is readily given by:
\begin{equation}
    E_x(t) = E_x(0) + \Gamma_0 R t \;.
\end{equation}
With no damping force from cooling ($\Delta=0$), the ion experiences linear heating at constant rate $\Gamma_0 R$.

\section{High $\bar{n}$ Measurement}
We estimate $\bar{n}$ for heating times up to 2 ms using a measured carrier Rabi oscillation. In these experiments, the ion is first prepared near its motional ground state. A carrier $\pi$-pulse then prepares a bright qubit state; the ion is then irradiated by the detection beam for a variable length of time, extending out to 2 ms.

Far-detuned 355 nm Raman beams couple to the $x$ and $y$ principal axes in this experiment, so both modes affect the carrier Rabi oscillation. A carrier Rabi oscillation with $x$ and $y$ COM mode couplings is given by
\begin{equation}
   P_{\text{bright}}(t) = \sum_{n_x, n_y=0}^{\infty} p_{n_x} p_{n_y} \sin(\Omega_{n_x, n_y} t / 2)^2
\end{equation}
where the Rabi frequency for the $x$ and $y$ mode couplings is given by
\begin{equation}
   \Omega_{n_x, n_y} = \Omega_0 e^{-\eta_x^2/2} e^{-\eta_y^2/2} \mathcal{L}_{n_x}(\eta_x^2) \mathcal{L}_{n_y}(\eta_y^2)
\end{equation}
with $\eta_x = 0.104$ and $\eta_y = 0.112$ measured independently \cite{wineland1998experimental}. The thermal contribution from the $y$-axis can be removed by approximating its contribution with the ratio of the respective secular frequencies $\bar{n}_y = (\omega_x / \omega_y) \bar{n}_x = 1.48 \bar{n}_x$. The motional distribution is assumed to be thermal. The final fitting parameters of the measured carrier Rabi oscillation are then the Rabi frequency and average motional state along the $x$-axis, $\{\Omega_0, \bar{n}_x\}$.

\end{appendices}
\end{document}